\documentclass[ussrhead]{elsart}

\usepackage{units}
\usepackage{epsfig}
\usepackage{amsfonts}
\usepackage{amssymb}

\journal{Astroparticle Physics}         

\def\Nmutr{N_\mu^{\rm tr}}
\def\Ne{N_{\rm e}}

\runauthor{Markus Roth}
\runtitle{A Non-Parametric Approach to Infer the Energy Spectrum and the Mass
  Composition of Cosmic Rays}

\begin{document}
\begin{frontmatter}
\title{A Non-Parametric Approach to Infer the Energy Spectrum and the Mass
  Composition of Cosmic Rays}
\author[KA-FZK]{T.~Antoni},
\author[KA-FZK]{W.\,D.~Apel},
\author[BU]{F.~Badea},
\author[KA-FZK]{K.~Bekk},
\author[KA-FZK]{K.~Bernl\"ohr\thanksref{nowatBerlin}},\relax 
   \thanks[nowatBerlin]{Now at:
      Humboldt Universit\"at, Berlin, Germany.}%
\author[KA-FZK,KA-Uni]{H.~Bl\"umer},
\author[KA-FZK]{E.~Bollmann},
\author[BU]{H.~Bozdog},
\author[BU]{I.\,M.~Brancus},
\author[KA-FZK]{C.~B\"uttner},
\author[YE]{A.~Chilingarian},
\author[KA-Uni]{K.~Daumiller},
\author[KA-FZK]{P.~Doll},
\author[KA-FZK]{J.~Engler},
\author[KA-FZK]{F.~Fe{\ss}ler},
\author[KA-FZK]{H.\,J.~Gils},
\author[KA-Uni]{R.~Glasstetter},
\author[KA-FZK]{R.~Haeusler},
\author[KA-FZK]{W.~Hafemann},
\author[KA-FZK]{A.~Haungs},
\author[KA-FZK]{D.~Heck},
\author[KA-Uni]{J.\,R.~H\"orandel},
\author[KA-FZK]{T.~Holst},
\author[KA-FZK,KA-Uni]{K.-H.~Kampert},
\author[LZ-Dep]{J.~Kempa\thanksref{nowatwarsaw}},\relax 
   \thanks[nowatwarsaw]{Now at: Technical University of Warsaw, Plock, Poland.}
\author[KA-FZK]{H.\,O.~Klages},
\author[KA-Uni]{J.~Knapp\thanksref{nowatLeeds}},\relax 
   \thanks[nowatLeeds]{Now at: University of Leeds, Leeds LS2~9JT, U.K.}%
\author[KA-FZK]{G.~Maier},
\author[KA-FZK]{H.\,J.~Mathes},
\author[KA-FZK]{H.\,J.~Mayer},
\author[KA-FZK]{J.~Milke},
\author[KA-FZK]{D.~M\"uhlenberg},
\author[KA-FZK]{M.~M\"uller},
\author[KA-FZK]{J.~Oehlschl\"ager},
\author[BU]{M.~Petcu},
\author[KA-FZK]{H.~Rebel},
\author[KA-FZK]{M.~Risse},
\author[KA-FZK]{M.~Roth\thanksref{corres}},\relax 
   \thanks[corres]{corresponding author; email: roth@ik3.fzk.de}
\author[KA-FZK]{G.~Schatz\thanksref{schatz_home}},\relax 
   \thanks[schatz_home]{Present adress: Habichtweg 4, D-76646 Bruchsal, 
                        Germany.}
\author[KA-FZK]{J.~Scholz},
\author[KA-FZK]{T.~Thouw},
\author[KA-FZK]{H.~Ulrich},
\author[YE]{A.~Vardanyan},                                                     
\author[BU]{B.~Vulpescu},
\author[KA-Uni]{J.\,H.~Weber},
\author[KA-FZK]{J.~Wentz},
\author[KA-FZK]{T.~Wiegert},
\author[KA-FZK]{J.~Wochele},
\author[LZ-Sol]{J.~Zabierowski}, 
\author[KA-FZK]{S.~Zagromski} 
\collab{(The KASCADE Collaboration)}
\address[KA-FZK]{Institut f\"ur Kernphysik, Forschungszentrum Karlsruhe,
             76021~Karlsruhe, Germany}
\address[BU]{National Institute of Physics and Nuclear Engineering,
             7690~Bucharest, Romania}
\address[KA-Uni]{Institut f\"ur Experimentelle Kernphysik, University of
             Karlsruhe, 76021~Karlsruhe, Germany}
\address[YE]{Cosmic Ray Division, Yerevan Physics Institute,
             Yerevan~36, Armenia}
\address[LZ-Dep]{Department of Experimental Physics,
             University of Lodz, 90236~Lodz, Poland}
\address[LZ-Sol]{Soltan Institute for Nuclear Studies,
             90950~Lodz, Poland}

\ifx AA
\makeatletter
\begingroup
  \global\newcount\c@sv@footnote
  \global\c@sv@footnote=\c@footnote     
  \output@glob@notes  
  \global\c@footnote=\c@sv@footnote     
  \global\t@glob@notes={}
\endgroup
\makeatother
\fi

\clearpage 
\begin{abstract}
The experiment KASCADE observes simultaneously the electron-photon,
muon, and hadron components of high-energy extensive  air showers
(EAS). 
The analysis of EAS observables for an
estimate of  energy and mass of the primary particle invokes extensive
Monte Carlo simulations of the EAS development for preparing
reference patterns. 
The present studies utilize the air shower
simulation code CORSIKA with the hadronic interaction
models VENUS, QGSJet and Sibyll, including simulations of
the detector response and efficiency. By applying non-parametric
techniques the measured data have
been analyzed in an event-by-event  mode and  the mass and energy of
the EAS inducing particles are reconstructed. Special emphasis is
given to methodical limitations and the dependence of the results on
the hadronic interaction model used.
The results obtained from KASCADE data reproduce the knee in the primary
spectrum, but reveal a strong model dependence. 
Owing to the systematic uncertainties introduced by the hadronic interaction 
models no strong change of chemical composition can be claimed in the energy 
range around the knee.

\end{abstract}
\end{frontmatter}

\clearpage 
\section{Introduction}
The basic astrophysical questions in 
high-energy cosmic rays (CR) relate to the sources,
the acceleration mechanisms and the propagation of CR through 
space. In particular, the observation of the change of the power law
slope (the {\em knee}~\cite{chrisitiansen}) of the all-particle spectrum at
an energy of a few times $\unit[10^{15}]{eV}$ has induced 
considerable interest and 
experimental activities. Nevertheless, despite of many conjectures and
attempts, the origin of the knee phenomenon has not yet been convincingly
explained.  
 
Due to the rapidly falling intensity and low fluxes, cosmic rays
of energies above $\unit[10^{14}]{eV}$ can be studied only indirectly by
observations of extensive air showers (EAS) which are produced by
successive interactions of the cosmic particles with nuclei of the
Earth's atmosphere.
EAS develop in the atmosphere as avalanche
processes in three different  main components: the most numerous
electromagnetic (electron-photon) component, the muon component and
the hadronic component. 
The properties of EAS are usually measured with large
ground-based detector arrays. In most experiments only one or two
components are studied. The KASCADE
experiment~\cite{doll90,kascade97} studies all three main components
simultaneously and a large number of shower parameters are registered for
each event. Their analysis to determine
the properties of the primary particle are obscured by the considerable
fluctuations of EAS development.  

\begin{figure}[t]
  \begin{center}
    \epsfig{file=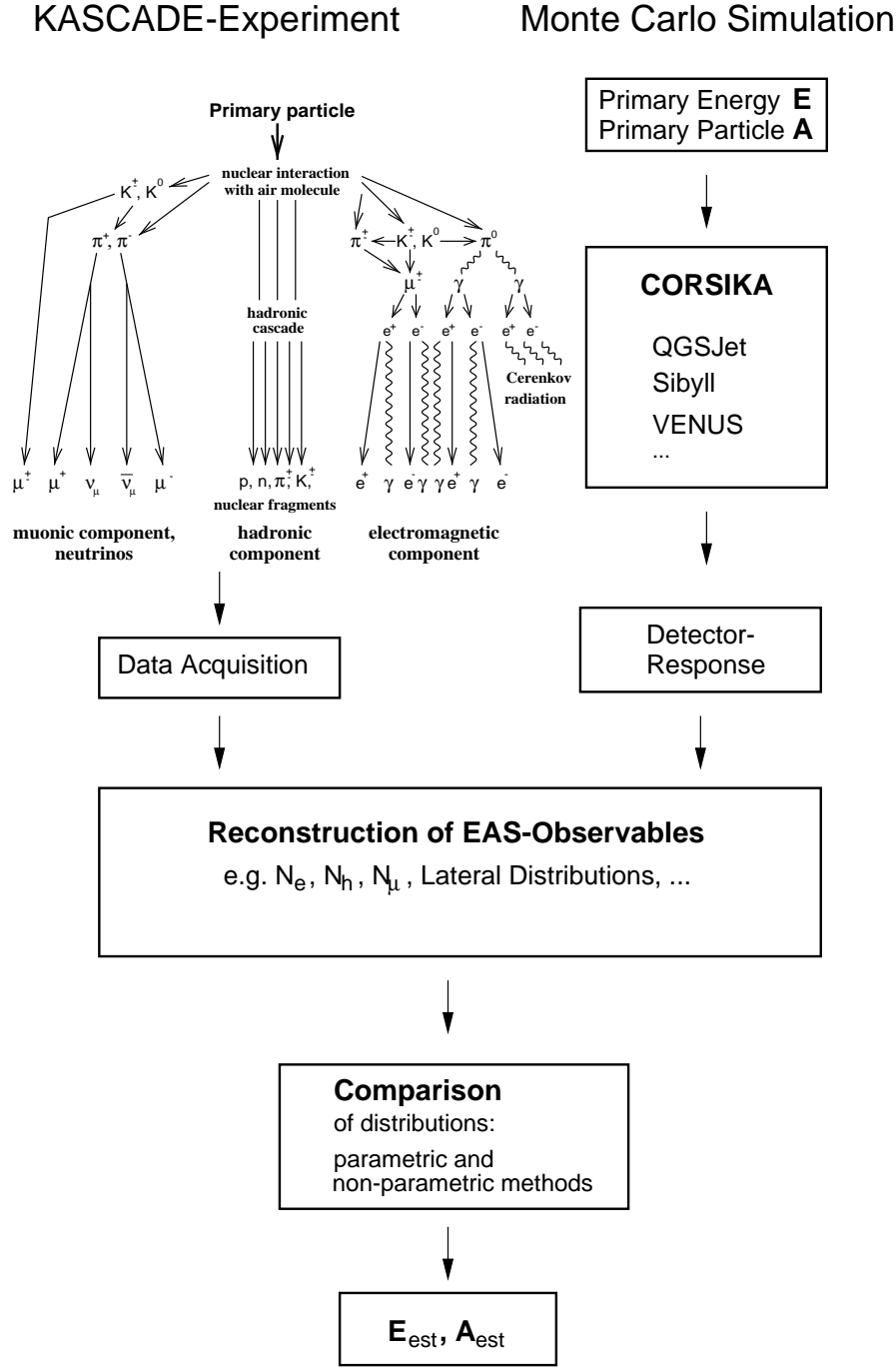,height=0.8\textheight}
  \end{center}
\caption {Twofold way of the EAS analysis procedure. }
\label{twofold:fig}
\end{figure}

The analysis of the EAS variables to deduce the properties of the 
primary particle relies on the comparison with Monte Carlo simulations (MC) of
the shower development (see Figure~\ref{twofold:fig}),  
including the detector response. Usually only one
or two EAS parameters are measured and various simplified
procedures are used to describe the relation between
the observed EAS properties and the nature and energy of the primary
particle. The simplification often implies the use of
parameterizations of the average behavior, which may bias
the results and limit the accuracy because fluctuations are neglected or not 
properly accounted for. For the 
analysis of multivariate parameter distributions 
and accounting for fluctuations more
sophisticated  methods are needed. The decades-old Bayesian methods and the
neural network approaches, currently in vogue, meet these necessities.  The
methods facilitate an event-by-event analysis. 

In the present paper we report on an investigation of the
energy spectrum and mass composition of cosmic rays in the energy range
of $\unit[10^{15}-10^{16}]{eV}$, based on the analysis of 700,000 EAS
events. 
A subset of approximately 8000 showers with 
cores  near the center of the hadron calorimeter yields information on all
three components and has been studied in more detail. Following
the analysis scheme shown in Figure~\ref{twofold:fig}, the simulated showers 
calculated with the simulation program CORSIKA~\cite{CORSIKA}  have been
convoluted with the apparatus response using the GEANT
code~\cite{geant93}. Non-parametric procedures~\cite{ashot89} yield not only
an estimate of the 
primary energy and mass composition, but they also allow to specify the
uncertainty of the results in a  quantitative way. 
In addition,  we specify the dependence of the results on the hadronic
interaction models. 
The necessity to invoke such models
in an energy range
beyond the experimental limits of
accelerator experiments, implies a model dependence of the results on the
energy spectrum and mass composition. Quantifying this model dependence is one
of the objectives of the present paper.
The model dependence is illustrated by using two different
interaction models for the analysis. The dependence implies
not only the degree to which a particular EAS observable is correlated to
energy and mass of the primary particle, it 
shows also how sensitively different EAS observables reveal primary mass.
As an example, the mass composition depends on the particular set of 
observables being considered simultaneously in the analysis if the model 
is inconsistent with the data in all internal correlations.

It should be stressed that the present study emphasizes the
methodical aspects of how to infer  energy spectrum and mass composition of CR
rather than providing a final answer. This would require  improved
statistical accuracy both in experiment and simulation and, first of all, a 
reduction of systematic uncertainties due to the incomplete knowledge of 
high energy interactions. Nevertheless our
findings on spectrum and mass composition are compatible within the methodical
accuracy to the results of other experiments.  

\section{The KASCADE experiment}
\begin{figure}[t]
  \begin{center}
    \epsfig{
       file={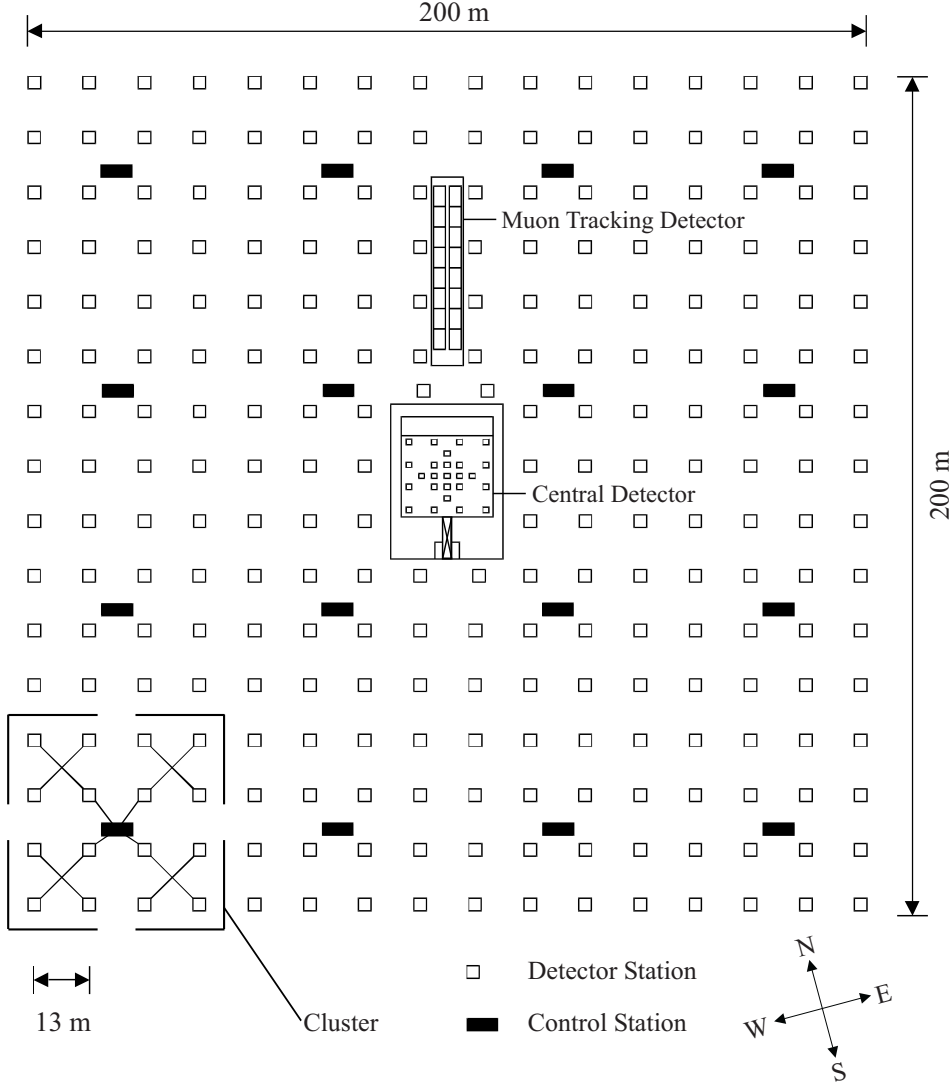},width=0.9\textwidth}
  \end{center}
\caption {Schematic layout of the KASCADE experiment. }
\label{layout:fig}
\end{figure}

The  detector installation of the experiment  KASCADE  (KArlsruhe  Shower 
Core  and Array  DEtector)~\cite{doll90,kascade97} is located on the site of 
the Forschungszentrum Karlsruhe, Germany ($8^\circ$\ E, $49^\circ$\ N;
\unit[110]{m} a.s.l.).  
The three major components of the detector system
(Figure~\ref{layout:fig}) are 
\begin{itemize}
\item an {\em Array} of scintillation detectors, 
\item a {\em Central Detector}: an arrangement of several different detector
  components, basically a hadron iron sampling calorimeter using liquid
  ionization chambers and 
\item a {\em Muon Tracking Detector} (MTD) using limited streamer tubes.
\end{itemize}
The Array covers an area of about $\unit[200\times200]{m^2}$ and
consists of 252 detector stations. These are organized in 16 clusters
and placed on a square grid of \unit[13]{m} separation.
The detector stations contain liquid scintillation
counters ($e/\gamma$ detectors) of $\unit[0.79]{m^2}$ area each and plastic
scintillators of $\unit[0.81]{m^2}$ each 
($\mu$ detectors; $E_\mu^{\rm thres}=\unit[230]{MeV}$), the latter covered by
a shielding of $\unit[10]{cm}$ lead and $\unit[4]{cm}$ steel. 
The inner four clusters (60 stations) contain four $e/\gamma$
detectors per station but no $\mu$ detectors while the outer 12
clusters (192 stations) have two $e/\gamma$ detectors and four $\mu$ detectors
per 
station. 
The reconstruction of the EAS data measured with the Array provides the basic 
information about lateral distributions and total intensities  of the
electron-photon (shower size $\Ne$)  and muon components ($\Nmutr$; see
Section~\ref{reconstruction:sec}), the location of the EAS core and the
direction of incidence. 

The layout and performance of the Central Detector are described in
reference~\cite{engler:cit}. 
The finely segmented hadron calorimeter is the main part of in the
Central Detector system.  It consists of a 
$\unit[20\times16]{m^2}$ stack of about
$\unit[4000]{tons}$ of iron with eight horizontal gaps. 
The calorimeter thickness corresponds to 11 interaction lengths
$\lambda_{\rm I}$ for vertical hadrons.
The detectors, measuring the energy deposit of the traversing 
charged particles, are  ionization chambers filled with the room temperature
liquids tetramethyl-silane  (TMS) or tetramethyl-pentane (TMP), inserted into
the gaps of the iron stack and read out by 40,000 electronic channels. From
their signals the impact point, the direction and the
energies of individual hadrons are reconstructed. 
In particular, the number of EAS hadrons with energies larger than
$\unit[100]{GeV}$  ($N_{\rm h}^{E>\unit[100]{GeV}}$),  the energy of the
most energetic hadron observed in the shower  ($E_{\rm h}^{\rm max}$) and the
energy sum of all reconstructed hadrons  ($\sum E_{\rm h}$)  are deduced as
shower observables (see Section~\ref{reconstruction:sec}). 

A layer  of 456 scintillation detectors, each with a size of
$\unit[0.45]{m^2}$, is mounted in the third gap at a depth of 2.2 $\lambda_{\rm
  I}$. It is used for triggering the Central Detector system, for muon 
detection
(with a threshold of $E_\mu^{\rm thres}=\unit[490]{MeV}$), and to determine 
arrival time distributions~\cite{timepaper00}.  

In the basement of the central building, below the iron stack and
\unit[77]{cm} of concrete, two layers of multi-wire proportional chambers
(MWPCs) are arranged as a tracking hodoscope, covering an area of
$\unit[122]{m^2}$~\cite{mwpcpaper00}. 
The MWPCs register muons with an energy threshold $E_\mu^{\rm thres}=
\unit[2.4]{GeV}$ and provide the observable $N_\mu^\star$, i.e. number of
reconstructed muons in the MWPCs. Due to the 
good position resolution, the MPWCs  register also the spatial
distribution of the high-energy muons together with traversing secondaries 
produced in the absorber by high-energetic hadrons, whose 
pattern has been shown to carry valuable information about
the mass of the primary particle~\cite{haungs96}, expressed by
particular parameters ($D_{-6}$, $D_6$) in terms of a fractal moment analysis
(see Section~\ref{reconstruction:sec}).  
\begin{table}[t]
\begin{center}
\caption{KASCADE detector components used in the present analysis.}
\label{detector:tab}
    \begin{tabular}[h]{llcr}
    Detector & Total Area & Threshold
    $E_{\rm{kin}}$ & Observables \\ 
    \hline
    array $e/\gamma$ & $\unit[490]{m^2}$ &
    \unit[5]{MeV} &  $N_{\rm{e}}$\\
    array $\mu$  & $\unit[622]{m^2}$ &
    $\unit[230]{MeV}\times \sec \theta$  &  $N_\mu$, $N_\mu^{\rm tr}$\\
    MWPCs  & $\unit[122]{m^2}$ &
    $\unit[2.4]{GeV}\times \sec \theta$ &  $N_\mu^\star$, $D_{-6}$, $D_{6}$\\
    calorimeter & $\unit[320]{m^2}$ &
    \unit[50]{GeV} & $N_{\rm{h}}^{E>\unit[100]{GeV}}$,
    $E_{\rm{h}}^{\max}$, $\sum E_{\rm{h}}$  \\

    \hline
    \end{tabular}
\end{center}
\end{table}
Information on  specific detector details is compiled  in Table 1.


\section{Simulations}
The simulations of the EAS development, along the requirements of the analysis
scheme of  Figure~\ref{twofold:fig}, have used the air
shower simulation program \mbox{CORSIKA} (vers.~5.62)~\cite{CORSIKA}. The  
code incorporates several options of
high-energy interaction models and is continuously under
improvement. In particular, we consider the latest
versions of VENUS~\cite{werner93}, QGSJet~\cite{kalmykov93} and Sibyll
(vers.~1.6)~\cite{engel92}.  
VENUS and QGSJet are models based on the
Gribov-Regge theory, and extrapolate the interaction features in a well
defined way into energy regions which are far beyond energies
available by accelerators, and especially into the extreme
forward direction. Sibyll is a minijet model used  as a hadronic interaction
generator in the MOCCA~\cite{hillas81} and the AIRES codes~\cite{aires}. We use
it here only for demonstration purposes. For 
the low-energy interactions \mbox{CORSIKA} includes the GHEISHA
code~\cite{gheisha}. The influence of the Earth magnetic field on
charged particle propagation is taken into account. As density profile of the
atmosphere the U.S. standard atmosphere is chosen~\cite{CORSIKA}.

Samples of at least 2000 proton and iron-induced showers have been simulated
with all three models.  Additionally for VENUS and QGSJet the
intermediate mass primaries He, O and Si have been simulated.
The energy distribution follows a weighted power law with a spectral index of
$-2.7$ in 
the energy range of $\unit[10^{14}]{eV}$ to $\unit[3.16\cdot10^{16}]{eV}$,
calculated  in eight intervals. The zenith angles are distributed in the range
$[13^\circ,22^\circ]$. The centers of  the 
showers  are  spread uniformly over an  area  which  exceeds the
surface of the  hadron  calorimeter  by \unit[2]{m} on each side. In addition, 
roughly the same number 
of simulated events with the centers of the showers within the Array are used.
The signals  observed in individual
detectors are determined by tracking all secondary particles down to 
observation  level  and passing them through  a  detector response
simulation program based on the GEANT package~\cite{geant93}.
  

\section{Event reconstruction and selection}
\label{reconstruction:sec}
The reconstruction of the EAS  observables which is described in detail in
preceding  publications of the KASCADE
collaboration~\cite{haungs96,lateral99,unger97,horandel98,weber99}, applies an
iterative procedure for reconstructing the  shower  size  parameters.
In a first step the shower core location is determined by a center-of-gravity
technique from the energy deposit signals of all $e/\gamma$ counters, and the
shower  direction  is estimated by  a  simple plane fit using the timing
information of the Array detectors. In addition, as rough first  
approximations, the  electron  size $\Ne$ and muon size $N_\mu$  are estimated
from summation of detector signals,  
taking into account the actual shower core position on the grid.
These parameter values  are initial values for the
further reconstruction steps. In the second step the shower direction is
determined by fitting a conical shape of the shower disc to the 
arrival times of the charged particle component, registered with the
$e/\gamma$ counters. The lateral distributions and their shape parameters are
estimated, and $\Nmutr$ and $\Ne$ are determined.

\begin{figure}[t]
  \centering
  \epsfig{file=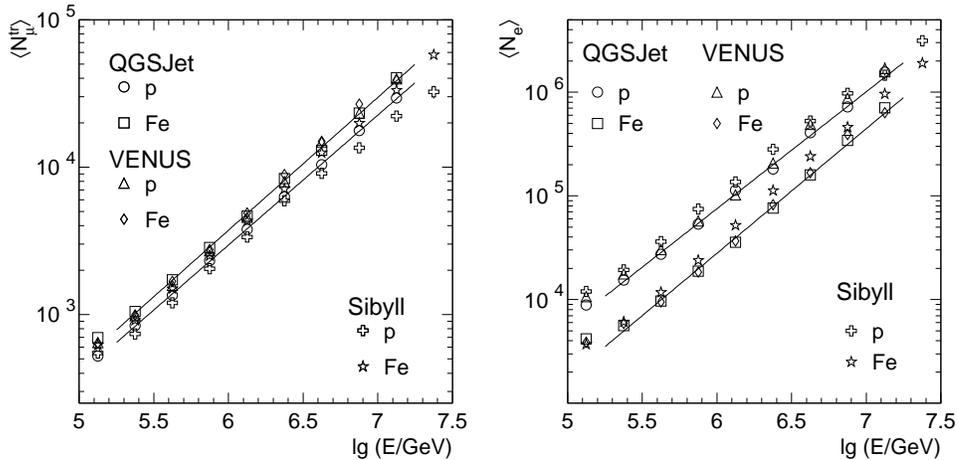,width=\textwidth}
  \caption{The mean values of the truncated muon number $\Nmutr$ and electron
    number $\Ne$ vs. primary energy as inferred on basis of the indicated
    interaction models. For sake of clarity only QGSJet predictions are fitted
    by a linear function in lg-lg scale, in order to emphasize the much more
    pronounced mass dependence of the shower size $\Ne$.} 
  \label{mcenenmu_paper:fig}
\end{figure}

\begin{figure}[t]
  \centering
  \epsfig{file=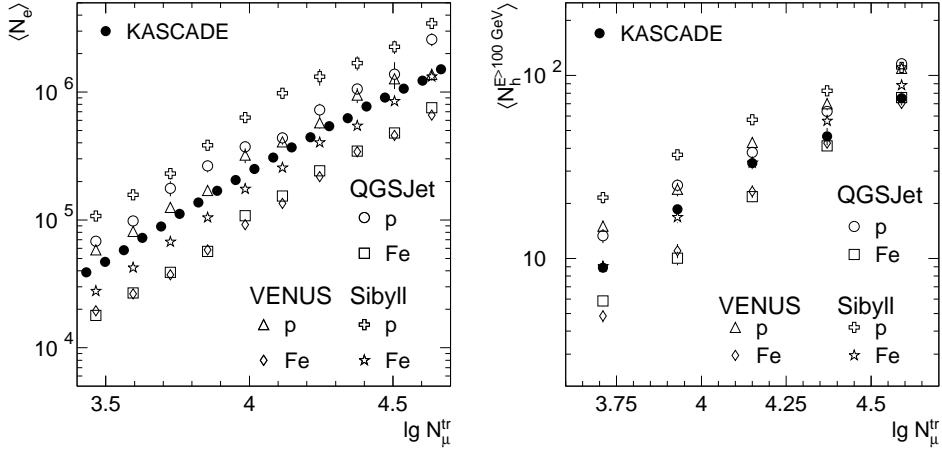,width=\textwidth}
  \caption{ The variation of the mean values of the electron shower  size
  (left) and number of reconstructed hadrons $N_{\rm h}^{E>\unit[100]{GeV}}$
    (right) with $\Nmutr$. The predictions of different models for proton
    and iron induced  showers are compared with results of the measurements.} 
  \label{nenmu_paper:fig}
\end{figure}

The muon size $\Nmutr$ is the  muon content within a  
range of distances from the shower core between \unit[40]{m} to
\unit[200]{m} which is the range  accessible to the KASCADE experiment 
(the so-called truncated muon number)~\cite{weber99}.  The lower 
limit is chosen to exclude contributions of  the 
electromagnetic and hadronic punch-through near the center of 
the showers. The upper limit corresponds to the geometrical acceptance of the
KASCADE layout. Figure~\ref{mcenenmu_paper:fig} displays the variation of
$\Nmutr$ and $\Ne$ versus the primary energy $E$, as inferred from EAS
simulations. Due to various serendipitous features of the KASCADE layout $\lg
\Nmutr$ proves to be nearly proportional to $\lg E$  and turns out to be
almost independent of primary mass (Figure~\ref{mcenenmu_paper:fig} left). 
This is in contrast to the electron
size $\Ne$, which exhibits  a strong mass dependence for fixed $\Nmutr$ as
shown in Figure~\ref{nenmu_paper:fig} (right).  

Contributions to the detector signals of other particles than electrons and
muons are eliminated by applying a lateral energy correction function to
appropriate particle densities, which are fitted with a likelihood function to
the Nishimura Kamata  
Greisen (NKG) formula~\cite{greisen56,kamata58}. Values of the radius
parameters of \unit[89]{m} and \unit[420]{m} 
for electrons and muons, respectively, are used~\cite{lateral99}.
For showers whose cores are located within \unit[91]{m}
from the Array center\footnote{
This number \unit[91]{m} results from the extension and the grid spacing of 
the detector array.}, the reconstruction uncertainty is about
\unit[2]{m} for the location of the shower center, $0.5 ^\circ$ for the angle
of incidence, and less than 10\% and 20\% for $\Ne$ and $\Nmutr$ values,
 respectively, at
primary energies larger than $\unit[10^{15}]{eV}$.

Muon tracks observed with the MWPCs, reconstructed from pairs of hits in the
two MWPC  
layers (vertically separated by \unit[38]{cm}~\cite{haungs96}), are summed up
to obtain $N_\mu^\star$. 
A limit for the reconstructed angle of $\pm 15^\circ$ in zenith and 
$\pm 45^\circ$ in azimuth with respect to the shower axis determined from the 
Array is imposed (the azimuth cut is not applied for showers with zenith 
angles of $<10^\circ$).
The analysis of the number and spatial distribution of the
muons and of produced secondaries in terms of two generalized multi-fractal
dimensions $D_{-6}$ and $D_{6}$ is discussed in~\cite{haungs96}. These
parameters characterize the spatial distribution of   
muons and high-energy (punch-through) hadrons as well as 
the degree of fluctuations of particles in the shower core.  

The reconstruction of  the hadronic shower variables applies appropriate
pattern recognition  
algorithms~\cite{unger97,horandel98}. Energy clusters found 
in different detector layers are traced from lower layers to the topmost one
 to form a particle track. Additionally, the angle of incidence of the
track can be deduced by the same procedure starting at lower layers, patterns  
of cascades have to form clusters from the remaining energy bunching up to
showers according to the already determined direction. Furthermore, signals
in the first layer from the top are not used for energy determination,
because their electromagnetic punch-through distorts the hadron signals. The
signals, weighted by the overlying absorber thickness are summed up   
and converted to hadron energies~\cite{engler:cit}. Similar to the shower size 
$\Ne$, the reconstructed number of hadrons ($N_{\rm h}^{E>\unit[100]{GeV}}$) 
exhibits a strong mass dependence for fixed $\Nmutr$ as well
(Figure~\ref{nenmu_paper:fig}; right). 

For the investigation of the primary energy spectrum and mass composition, as
well as of particular correlations of observables, two sets of data are
compiled.  
One set - further referred to as {\em selection I\/} - uses the information of
electrons and muons from the Array stations only. It allows analysis the
data with good statistical accuracy, but includes only little  information
provided by the Central Detector. A second data set - henceforth referred to as
{\em selection  II\/} - includes observables measured in the Central Detector,
but this data sample comprises only a small amount of registered
showers. {\em Selection~I\/} comprises 720,000 EAS events, accumulated in 226
days, with  primary energies 
larger than $E\approx 5\cdot\unit[10^{14}]{eV}$  and with a maximum core  
distance from the center of the Array of \unit[91]{m} and with
angles of incidence in  the range of $[13^\circ,22^\circ]$.      
{\em Selection~II} comprises approximately 8000 high-energy, central
showers selected by cuts on $\Nmutr$ ($\lg \Nmutr > 3.5$), on the core
location ($R_{\rm core}<  \unit[5]{m}$ from the center of the Central 
Detector), with at least one reconstructed hadron of an energy above
\unit[100]{GeV} and 10 muons observed in the MWPCs.  

\section{Non-parametric analyses}
The present analysis of mass
composition and energy spectrum avoids the
bias inherent in parametric procedures and is performed for individual
events by use of multivariate non-parametric Bayesian and neural
network decision methods. In this way we are  able to specify, in a transparent
and coherent way, how conclusive and trustworthy our results are, as expressed
by true classification and misclassification matrices of the results. A brief
outline and more details of the applied methods are given in 
Appendix~\ref{sec:methods} and in reference~\cite{chiliguide}.

The combination of the total muon
content $N_\mu$ and the shower size $\Ne$ has been shown to be sensitive to 
primary mass and is applied in numerous experimental studies, using suitable 
parameterizations of the predicted $\lg N_\mu / \lg \Ne $ relation with the
primary mass. 
However, as indicated above, the total muon content $N_\mu$,
although displaying some dependence on primary mass, is a quantity not
easily accessible experimentally without additional assumptions about the shape
of the lateral muon density distribution at large distances from the shower
core. Therefore,  we prefer to consider the truncated muon number $\Nmutr$,
which - on average - proves to be nearly independent from primary mass
(see Figure~\ref{mcenenmu_paper:fig}), but it is, on the other hand, a rather
sensitive energy identifier. Thus, at fixed $\Nmutr$,
the information about the mass is essentially provided by the shower size
$\Ne$~\cite{weber99}. In cases of other EAS  observables mass and energy
sensitivities are,  in general, less well marked, and in principle, each
shower variable carries information simultaneously on  mass and energy in
a way which is additionally affected by the considerable fluctuations of the
shower development. The most sensitive EAS observables,
$\Ne$ and $\Nmutr$, display the smallest intrinsic and sampling fluctuations.  


\subsection{Mass composition}
Due to the limited number of simulated EAS and the correspondingly limited
statistical accuracy it is hardly reasonable to use the full set of
observables simultaneously to achieve a reliable result about  mass
composition (curse of dimensionality condition; see
Appendix~\ref{sec:methods}). Hence we consider simultaneously only a few
observables.

Each simulated or measured event is represented by an observation vector 
$x=(\Ne,\Nmutr,\ldots)$ of the $n$ observables. 
Applying the technique described in  Appendix~\ref{sec:methods} the
likelihood   
(probability density distribution) $\hat{p}(x|\omega_i)$ of an event for each
class  $\omega_i\in\{ {\rm p,O,Fe} \}$ can be calculated,  i.e. the 
probability of an event $x$ belonging to a given class $\omega_i$.

\begin{figure}[t]
  \begin{minipage}[b]{0.5\linewidth}
    \epsfig{file=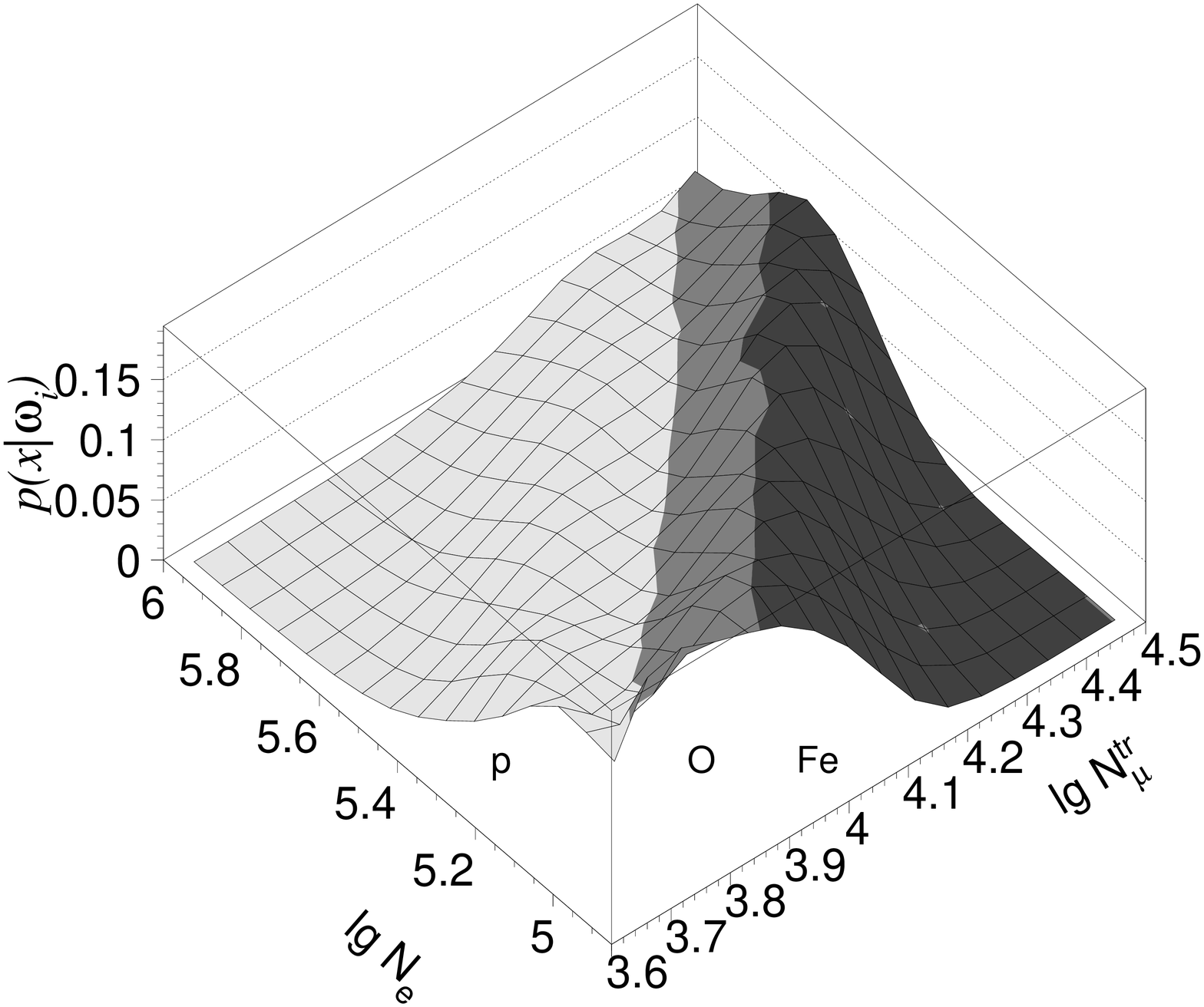,width=\textwidth}
  \end{minipage}
  \begin{minipage}[b]{0.5\linewidth}
    \epsfig{file=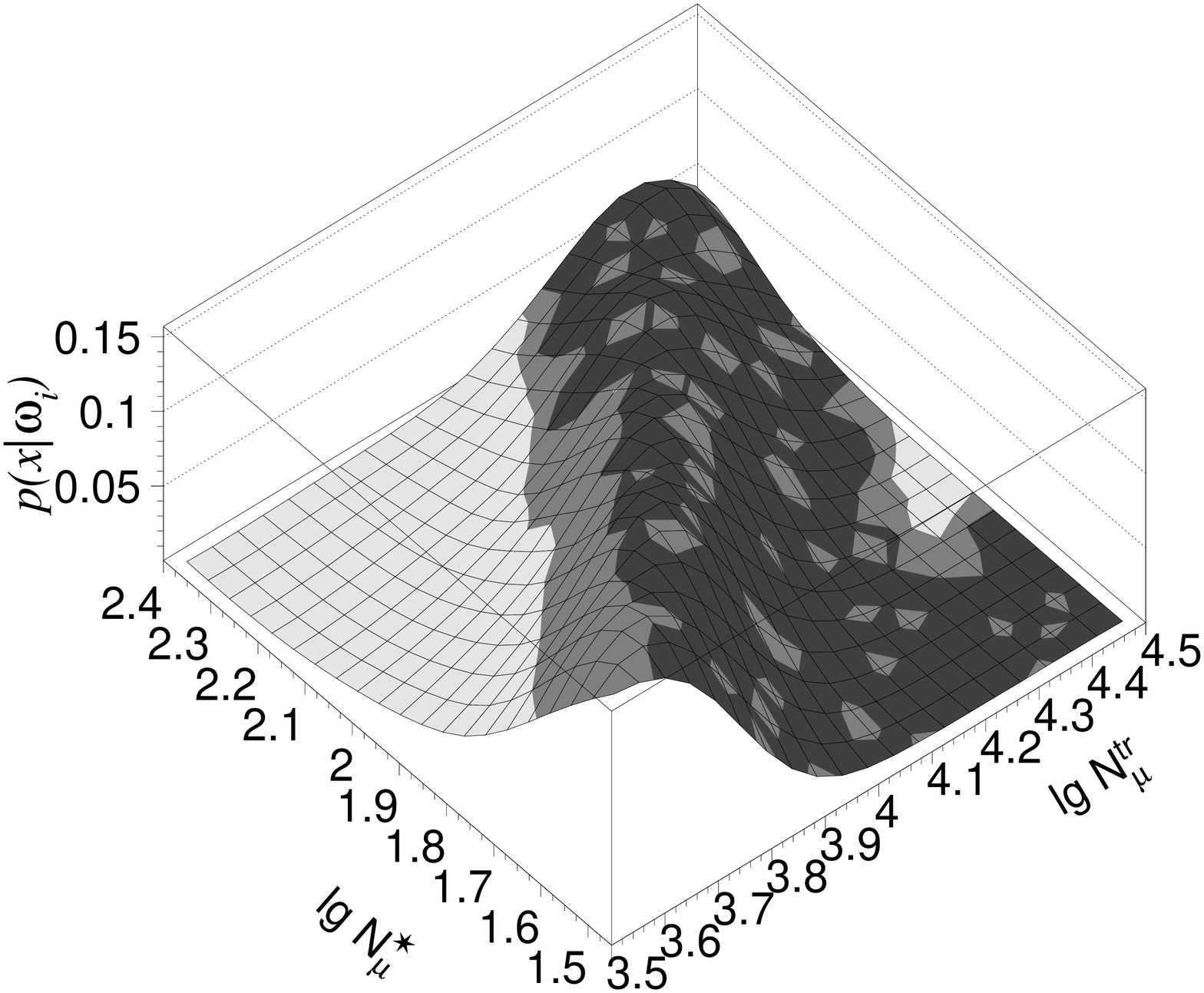,width=\textwidth}
  \end{minipage}
  \vspace*{-1.5cm}
  \caption{Superposition of three probability density distributions
    $\sum_{i=1}^3 \hat{p}(x|\omega_i)/3$ deduced
           from QGSJet simulations using  
           the observables $\Ne$ and $\Nmutr$ (left) as well as 
           $N_\mu^\star$ and $\Nmutr$ (right). Events in the dark shaded 
           area mark the region classified as iron, middle grey as oxygen, and 
           light grey as proton ({\em selection~II}).}
           \label{p_density:fig}
\end{figure}

As an example, the superposition of the 
estimated probability density distributions,
referring to two sets of different observables, are displayed in 
Figure~\ref{p_density:fig} (based on 
QGSJet simulations).
The regions where $\hat{p}(x|\omega_{\rm p})$, $\hat{p}(x|\omega_{\rm O})$ and
$\hat{p}(x|\omega_{\rm Fe})$ are larger than the other two possibilities are
colored light, middle and dark grey, respectively.
The left graph shows the density distribution calculated in the two 
dimensional space of the observables $\Ne$ and $\Nmutr$. 
A rough separation can be recognized, but also a strong overlapping  of the
likelihood distributions has to be admitted.

The right-hand graph of Figure 5 shows an example of two observables
($N_\mu^\star$ and $\Nmutr$), which exhibit only weak mass-discrimination
power. Correspondingly, the density distributions of the three particle types
are intermixed, and reliable conclusions could not be drawn.  
In case of  {\em selection~II} the mass composition is reconstructed for
different sets of observables using the Bayes theorem
(Equation~\ref{eqn:bayestheorem}).  
When the estimated {\em posterior \/} probability
$\hat{p}(\omega_i|x)$  is larger than $\hat{p}(\omega_j|x)$, then 
the event is assigned to class $\omega_i$, otherwise to class
$\omega_j$. Taking into account (by  
Equation~\ref{eqn:bayestheorem}) the estimated number of incorrectly
classified events (i.e. misclassification rates)
(Table~\ref{tab:34class3}) the true proportions of the different particle
types are reconstructed.  
  
The classification rates  
$P_{ij}=\hat{P}_{\omega_i\rightarrow\omega_j}$ (see Appendix:
Equation~\ref{sec:bayes} and Figure~\ref{callgaussbayes_paper:fig}) give the
fraction of correctly, $P_{ii}$, and wrongly, $P_{ij}$, classified events with
$i\neq j$, an example for three mass class is given in
Table~\ref{tab:34class3}. Of course, the sum of each row has to be 100\%. 
\begin{table}[t!]
\centering
\caption{Classification matrices for three classes (p, O and Fe) and two 
         different models. In addition to the classification rates of p, O
         and Fe, the rates of classified intermediate groups He and Si
         respectively, are given. The used observables are $\Nmutr$ and 
         $\Ne$  ($3.6\le\lg \Nmutr< 3.9$).}
\begin{tabular}{lrrrrrr}
            & \multicolumn{3}{c}{QGSJet} & \multicolumn{3}{c}{VENUS} \\[0.5ex]
\hline
$\hat{P}_{\omega_j\rightarrow \omega_i}$~[\%]& $\omega_i=$p  & $\omega_i=$O    & $\omega_i=$Fe   &\rule{18mm}{0mm} $\omega_i=$p       & $\omega_i=$O    & $\omega_i=$Fe\\[0.5ex]
\hline
$\omega_j=$p       & $ 77 \pm 3 $ & $ 21 \pm 3 $ & $ 2 \pm 1 $   & $78 \pm 3 $ & $21 \pm 2 $ & $ 1^{+2}_{-1} $ \\
$\omega_j=$He      & $ 57 \pm 3 $ & $ 39 \pm 3 $ & $ 4 \pm 1 $   & $64 \pm 3 $ & $32 \pm 2 $ & $4  \pm 1 $ \\
$\omega_j=$O       & $ 14 \pm 2 $ & $ 61 \pm 3 $ & $25 \pm 3 $   & $15 \pm 2 $ & $61 \pm 4 $ & $24 \pm 3 $ \\
$\omega_j=$Si      & $  3 \pm 1 $ & $ 54 \pm 3 $ & $43 \pm 2 $   & $ 3 \pm 2 $ & $51 \pm 3 $ & $46 \pm 2 $ \\
$\omega_j=$Fe      & $  1 \pm 1 $ & $ 17 \pm 2 $ & $82 \pm 3 $   & $ 0 ^{+1}_{-0} $ & $20 \pm 3 $ & $80 \pm 3 $\\
\hline
\end{tabular}
\label{tab:34class3}
\vspace*{0.5cm}
\end{table}

In the most probable cases the
different particle types are identified correctly, but the knowledge 
of the incorrectly classified events could be used for a correction due to the 
mis-classification.
In addition, the rates for the intermediate mass particle   
types, He and Si, are given. Helium is mostly classified as protons (57\%) and
silicon as oxygen (54\%). Due to the stronger fluctuations and weaker
correlations with mass and/or energy, different sets of observables
result in lower true-classification rates $P_{ii}$. In general, if  the rates
$P_{ii}$ are less than 50\%, it is no more possible to deconvolute the true
proportions by matrix inversion of $P_{ij}$ (Equation~\ref{eqn:reconstr}),
since the matrix $P_{ij}$ becomes singular, signaling  that the determination
of a class $\omega_i$ is just haphazard. Therefore it is not meaningful to
consider more than 3 classes, since this would require an analysis of further
observables simultaneously, with a number of Monte Carlo simulations larger
than presently available.   

\begin{figure}[t]
  \begin{minipage}[b]{0.5\linewidth}
    \epsfig{file=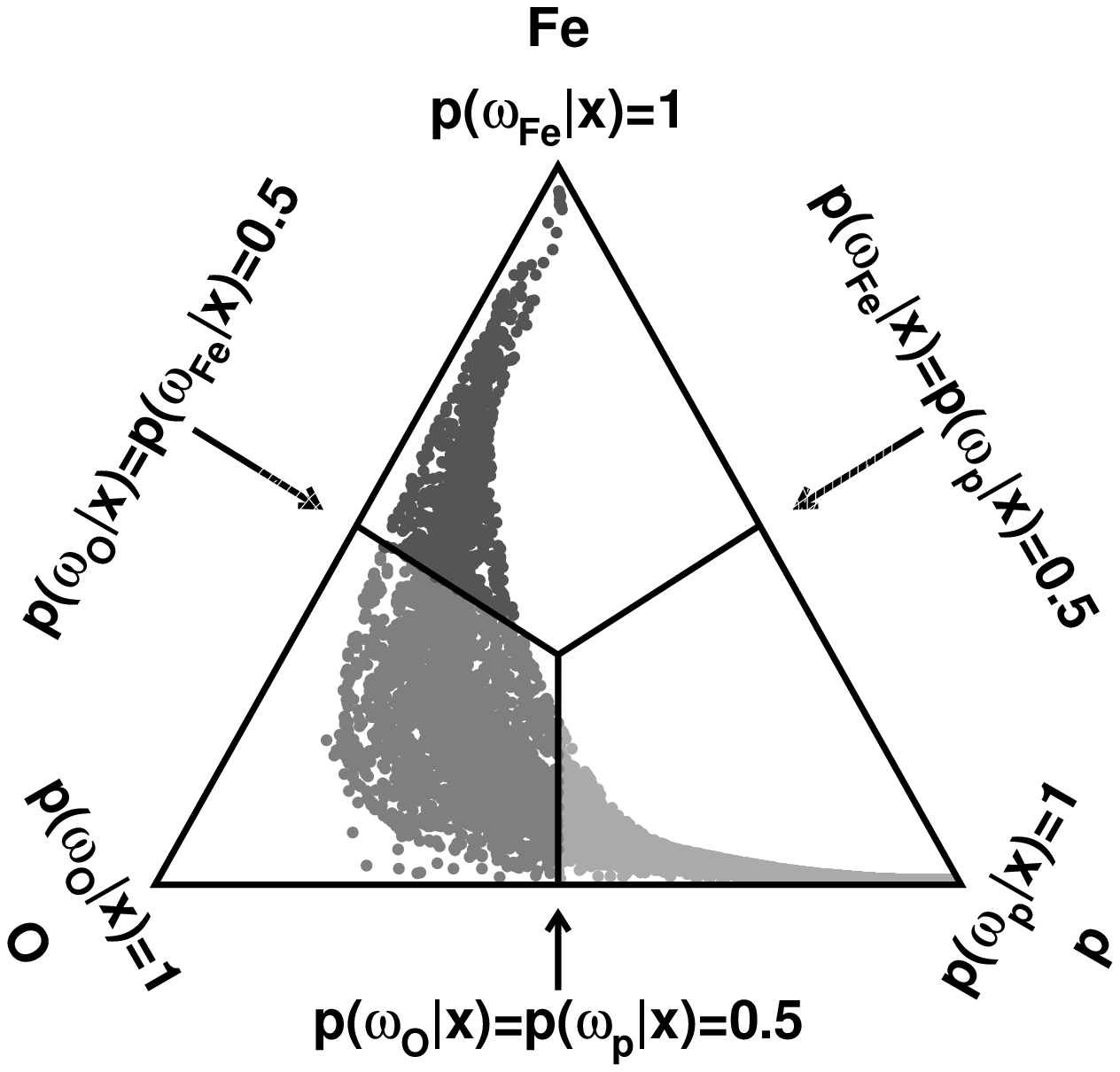,width=\textwidth}
  \end{minipage}
  \begin{minipage}[b]{0.5\linewidth}
    \epsfig{file=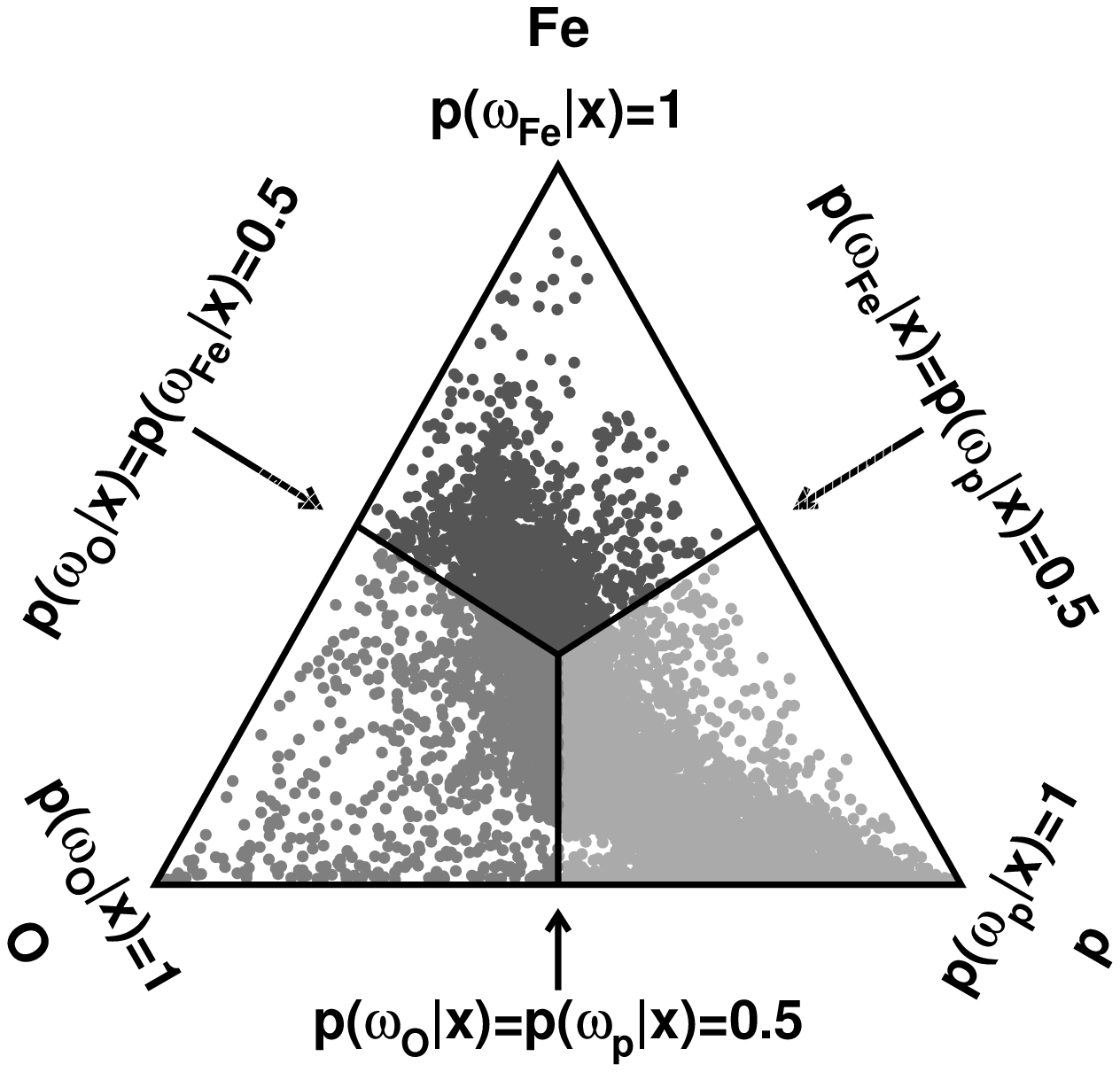,width=\textwidth}
  \end{minipage}
  \vspace{-2.cm}
  \caption{Estimated posterior probabilities  $\hat{p}(\omega_i|x)$ 
           of measured events
           deduced from QGSJet simulations using 
           the observation vector $x=(\Ne, \Nmutr)$ (left) as well as 
           $x=(N_\mu^\star, N_{\rm{h}}^{E>\unit[100]{GeV}},
           \sum E_{\rm{h}})$ (right).}
           \label{posterior:fig}
  \end{figure}
As a cross-check the estimated {\it posterior} probabilities 
$\hat{p}(\omega_i|x)$ of a given measured 
event $x$, belonging to class $\omega_i$, can be calculated 
(Figure~\ref{posterior:fig}). The center of the triangles shown
correspond to equal probability of belonging to any class, reflecting the 
fact that it is nearly impossible to classify the 
measured event, while points in the corners satisfy the relation 
$\hat{p}(\omega_i|x)=1$, i.e. the corresponding 
event belongs to class $\omega_i$ with probability unity.
Hence, from given measured events we obtain information about the 
probabilities belonging to class $\omega_i$ just as well.
Evidently, the set $\Ne$ and $\Nmutr$ allows to determine a well defined 
mass composition. In contrast, the set comprising $N_\mu^\star$, 
$N_{\rm{h}}^{E>\unit[100]{GeV}}$, and $\sum E_{\rm{h}}$ is less suitable for 
mass discrimination mainly due to strong fluctuations of the observables as
can be seen in Figure~\ref{posterior:fig}, right side.

\begin{figure}[t]
\epsfig{file=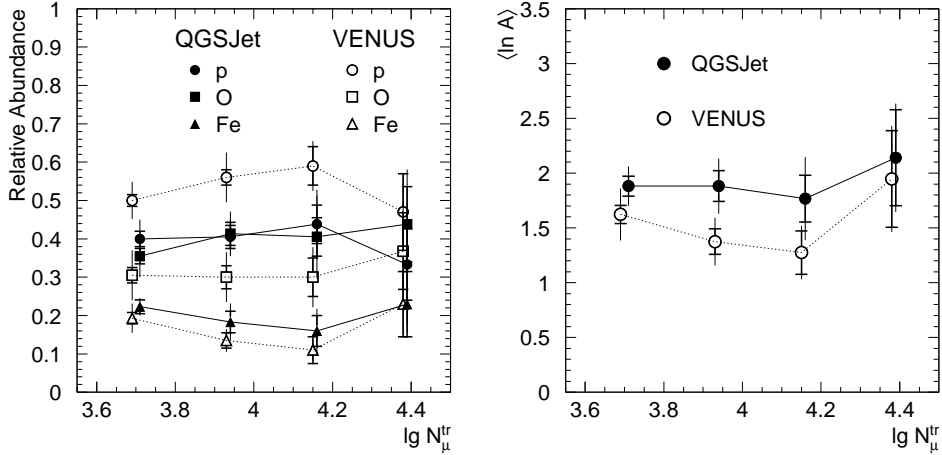,width=\textwidth}
  \caption{The relative abundances of the classes p,
    O and Fe vs. $\lg \Nmutr$, reconstructed on the basis of two different
    hadronic interaction models and using the EAS observables $\Ne$ and
    $\Nmutr$. The right graph shows the corresponding mean logarithmic mass
    $\langle\ln A\rangle$ vs. $\lg \Nmutr$. Statistical (thick lines) and
    methodical  (thin lines) uncertainties are indicated as error bars.}
           \label{comp2mean:fig}
\end{figure}
\begin{figure}[t]
  \epsfig{file=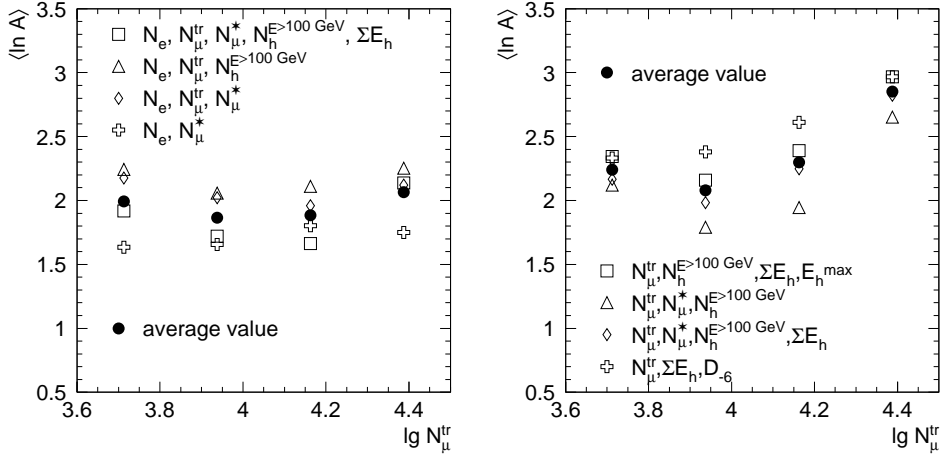,width=\textwidth}
  \caption{Mean logarithmic mass $\langle\ln A\rangle$ resulting from the
    analysis of different sets of observables vs. $\lg \Nmutr$ (QGSJet
    prediction). The sets displayed on the right do
    not include the observable $\Ne$. The error bars are omitted
    to simplify 
    the presentation of the synopsis. The statistical errors are of the same
    order of magnitude as given in Figure~\ref{comp2mean:fig}. But the
    systematic errors are larger by $40\%-50\%$ for the data on the right
    graph due to a weaker correlation with mass. }  
           \label{comptest:fig}
\end{figure}
The results of the composition determination, using the observables
$\Ne$ and $\Nmutr$, are given in Figure~\ref{comp2mean:fig} as 
relative abundances versus $\Nmutr$ and mean logarithmic mass
$\langle\ln A\rangle$ vs.  $\Nmutr$. 
The mean
logarithmic mass $\langle\ln A\rangle$ 
cannot be calculated unambiguously because only abundances of groups of 
elements can be determined. We have assigned 
$\langle  
A_{\rm p+He}\rangle=2.5$, $\langle A_{\rm O}\rangle=16$, 
$\langle A_{\rm Si+Fe}\rangle=42$ to the p, O and Fe group, respectively.
This procedure is of course to some extent arbitrary, but this is 
always implicit, when $\langle\ln A\rangle$ is used.
The statistical errors (thick lines) are calculated according to 
a multinomial distribution. The thin error bars correspond to a methodical
uncertainty, calculated by the bootstrap method (see 
Appendix~\ref{sec:methods}). It reflects the influence of the limited number
of Monte Carlo events. Apparently the use of the VENUS model 
results in a lighter composition, compared with the use of QGSJet.
A tendency towards a heavier composition above the knee 
($\lg \Nmutr=4.15$ corresponding to $\unit[4]{PeV}$) is indicated,
albeit the statistical and systematic uncertainties do not allow  a
definite conclusion and the results are clearly compatible with an energy
independent composition. 

Results of several other combinations of observables by using the QGSJet
model are summarized in Figure~\ref{comptest:fig}. In general, the tendencies
are the same. Remarkably, all sets omitting the electron size $\Ne$
(right graph) result in a heavier composition and a more pronounced increase
above the knee. 
As the electron size has the strongest mass sensitivity, as
well as the smallest fluctuations, the mass compositions are predominantly
determined by $\Ne$ and $\Nmutr$ (left). Compositions resulting from sets
of less sensitive observables differ from these values (right). 
The tendencies are quite similar for the VENUS model, but the 
absolute values are shifted towards a lighter composition
as expected from Figure~\ref{comp2mean:fig}. 

The fact that different combinations of observables taken into account in the 
analysis, lead  apparently to different mass compositions (shown in 
Figure~\ref{comptest:fig}), reveals inadequacies of the reference model, 
i.e.\ that the degree of the intrinsic correlations of different observables 
differs from those of the real data. Otherwise the determined mass 
compositions should be identical within the statistical errrors.


\subsection{Energy spectra}
To estimate the primary energy $E$ the most important parameters are
 $\Ne$ and $\Nmutr$ again, where
now $\Nmutr$ carries most of the information. As data basis we use {\em 
selection I}.
Due to the large computing time requirements we do not apply the Bayesian
algorithms here and use instead neural networks only. In principal there are 
no basic arguments to prefer one particular method. Previous publications have
 demonstrated the consistency and equivalence of neural network and Bayesian
 methods in EAS analyzes~\cite{roth99,ashot97}.  
The neural networks employed have typically a simple net topology $2\times 5
\times 2 \times 1$, but several other topologies have been used to estimate
the methodical error of this special choice. For training the network
(according to Equation~\ref{eqn:nngoal}) two independent samples have 
been generated to allow a  validation of the results
(Appendix~\ref{sec:methods}).  
\begin{figure}[t]
  \centering
  \epsfig{file=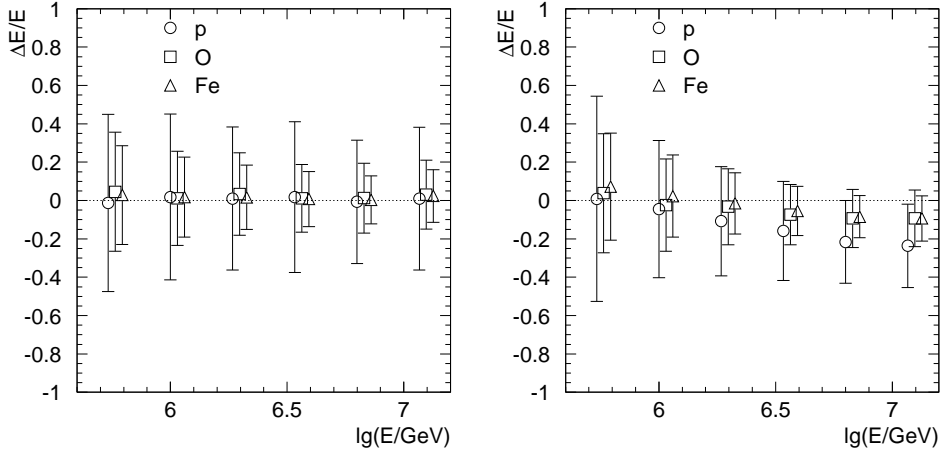,width=\textwidth}
  \caption{The relative error vs. primary energy for different classes:
  Results obtained with a network trained with QGSJet samples (left)  and  the
    result of the same network analyzing VENUS samples
    (right).}  
  \label{mlp_e_test_paper:fig}
\end{figure}
\begin{figure}[t]
  \centering
  \epsfig{file=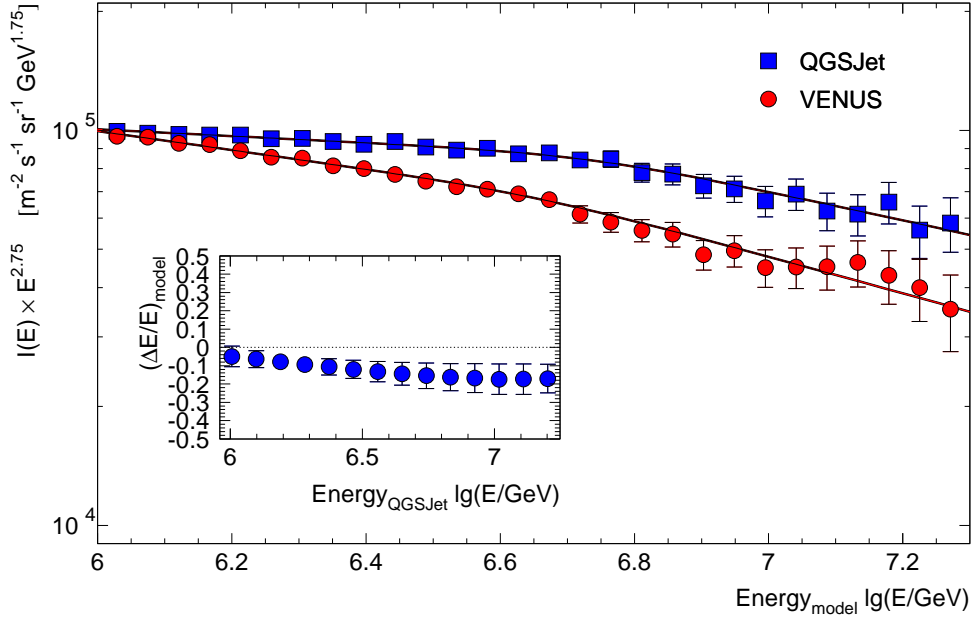,width=\textwidth}
  \caption{Differential energy spectra resulting from  the analysis of data of
  the KASCADE 
  experiment using two  differently trained  networks (by VENUS  and QGSJet
    samples). The  reconstructed energies are compared on an event-by-event
    basis and their differences are given in the inset as relative error
    vs. the energy reconstructed on the basis of the QGSJet model.} 
  \label{mlp_energy13-22_paper:fig}
\end{figure}
Before data analysis, the response and the biases of the trained neural
network have to be carefully scrutinized. 
For this, the performance of mass classification and energy estimation has
been investigated with the other MC sample as input.
We consider the relative deviation of the 
reconstructed energy $E_{\rm est}$ from the true value $E_{\rm true}$, which
is known for the simulated samples, more precisely, the distribution of
$\Delta E/E = (E_{\rm est} - E_{\rm true})/ 
E_{\rm true}$, whose mean value and the standard deviation represent 
the bias and the energy
resolution (relative error) of the reconstruction, respectively. 

Figure~\ref{mlp_e_test_paper:fig} displays the relative error of the estimated
energy for different primary particles. The  
relative error of a network, trained with QGSJet samples, is shown in the left
part. In general, the bias for the various classes is less than 3-5\%, but
the energy resolution (spread) proves to be strongly mass dependent. As
expected, the iron class has the smallest energy spread
($\sigma_{Fe}\approx21\%$ versus $\sigma_{p}\approx38\%$). The network trained 
with VENUS samples leads to the same results. Also shown (right) are the
results of a network, which has been trained with QGSJet samples analyzing
events generated with the VENUS model. With increasing energy the QGSJet
trained network underestimates systematically the true energy of the VENUS
samples. Moreover, the bias appears to be mass dependent, implying that the
degree of the correlation between energy and the analyzed shower observables
is varying differently for the different primary particles and models. This is
a caveat to the estimate of true energies  of the measured event samples, 
namly that hidden mass dependent correlations lead to an all-particle spectrum 
depending on the true  mass composition. 

Figure~\ref{mlp_energy13-22_paper:fig} presents the reconstructed energy
spectra of measured data resulting from the analysis using two different
networks, trained with QGSJet and VENUS samples. Apparently, the VENUS 
trained network results in a steeper spectrum as compared with
the QGSJet findings. It should be emphasized that the network used takes into
account not only the absolute values of the observables $\Ne$ and $\Nmutr$,
but also their correlations. In order to specify the relative error
arising from the model dependence, mean value and spread of $\left(\Delta E /E 
\right)_{model}=
(E_{\rm VENUS}-E_{\rm QGSJet})/E_{\rm QGSJet}$  are additionally given
(inset). The variation of this model error displays a change at higher
energies, which might indicate a change of the composition.

The resulting spectra are fitted by the trial function (see
Figure~\ref{mlp_energy13-22_paper:fig}) 
\begin{equation}
I(E)=I_0\cdot \left(\frac{E}{E_{\rm knee}}\right)^{-\gamma_1} \cdot 
        \left(1+\left(\frac{E}{E_{\rm knee}}\right)^\varepsilon\right)
                ^{\frac{\gamma_1-\gamma_2}{\varepsilon}}
\end{equation}
\begin{table}[t!]
  \centering
  \caption{Comparison of parameters of the energy spectrum, derived on basis
    of on the  VENUS and QGSJet simulations, respectively. The first error
    is the statistical one. The second error represents the systematic
     uncertainty resulting from the  small number of simulated event.} 
\begin{tabular}{ll@{$\pm$}l@{$\pm$}ll@{$\pm$}l@{$\pm$}l}
            & \multicolumn{3}{c}{QGSJet} & \multicolumn{3}{c}{VENUS} \\[0.5ex]
\hline
 $\gamma_1$ & 2.77 & 0.003 & 0.03 & 2.87 & 0.003 & 0.04\\
 $\gamma_2$ & 3.11 & 0.02  & 0.06 & 3.25 & 0.02 & 0.06\\
 $E_{\rm knee}$ [$\unit[10^{6}]{GeV}$] & 5.5 & 0.2 & 0.8 & 4.5 & 0.3 & 0.9 \\
  $\chi^2/dof$  & \multicolumn{3}{c}{0.95} & \multicolumn{3}{c}{1.94} \\
\hline
\end{tabular}
\label{tab:fitresults}
\end{table}
which accounts for a smoothly changing power law spectrum~\cite{samvel}. The
parameter $\varepsilon$ controls the width of the transition region, and the
\begin{figure}[t]
  \centering
\epsfig{file=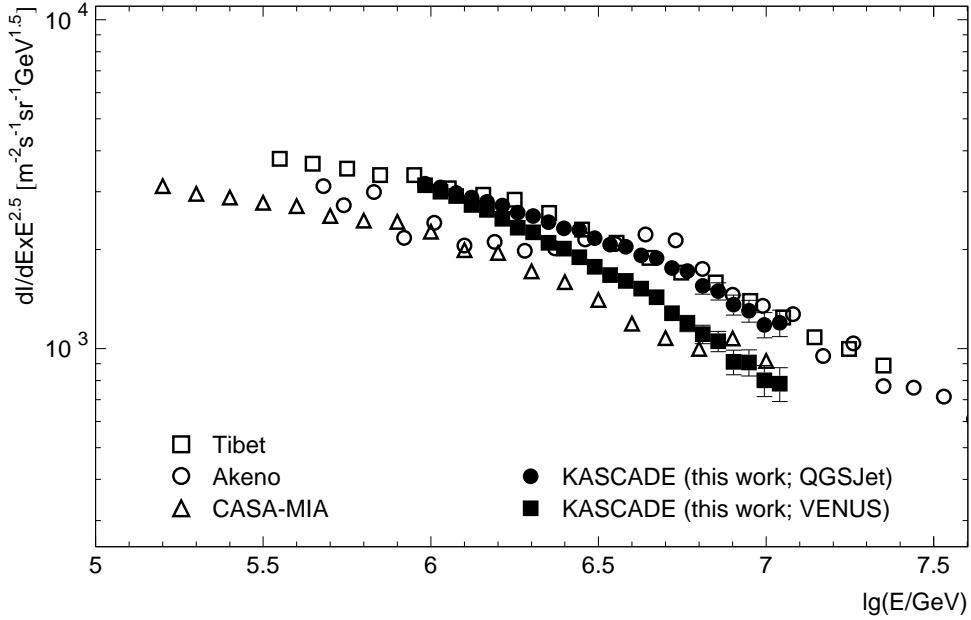,width=\textwidth}
  \caption{The differential all-particle cosmic ray flux obtained here 
           compared to the results reported by 
           Tibet AS$\gamma$~\cite{tibet}, Akeno~\cite{akeno}, and
           CASA-MIA~\cite{casa-mia}.
           The data points are multiplied by $(E/GeV)^{2.5}$. 
           Only statistical errors are presented.}  
  \label{othexp_paper:fig}
\end{figure}
knee position $E_{\rm knee}$ is defined by the center point of the transition
region. Asymptotically $I(E)$ approaches to power law functions
$E^{-\gamma_1,\gamma_2}$. The parameter uncertainties have  been studied by
calculating the errors $I(E)\pm \Delta I(E)$ using the sampling correlation
matrix. But the resulting error bands are so narrow that it  does not visibly
differ from the $I(E)$-line. The best-fit results are given in
Table~\ref{tab:fitresults}, including statistical errors as well as the
methodical error derived from different training parameters of the neural
network. It is obvious that the statistical errors are considerably smaller
than the systematic uncertainties resulting from the small number of simulated
events and from interaction models.

Figure~\ref{othexp_paper:fig} compares the spectra of 
Figure~\ref{mlp_energy13-22_paper:fig} with results reported by
other experiments. 
All measurements, independently from each other, show a steepening above a
particular energy: the knee. But the absolute intensity of the flux and the
position of the knee obviously differ. This is most likely due to different
model assumptions or energy conversion functions used. The considerable
deviation between CASA-MIA~\cite{casa-mia} 
and the other two experiments may be explained in this way. CASA-MIA used the
Sibyll model for 
constructing an energy estimator $E = f(\Ne , N_\mu)$ from the electron and
muon sizes. The fact that Sybill predicts significantly lower values of
$N_\mu$ and larger values for $\Ne$~\cite{antoni99} as compared to QGSJet and
VENUS, could lead to a systematic shift of the spectrum towards lower
energies.  
In view of the considerable model dependence of our results, the overlap with
some of the other experiments should not be taken as evidence for or against
any of them.


\subsection{Combined analysis of energy and mass}
\begin{figure}[t]
  \centering
  \epsfig{file=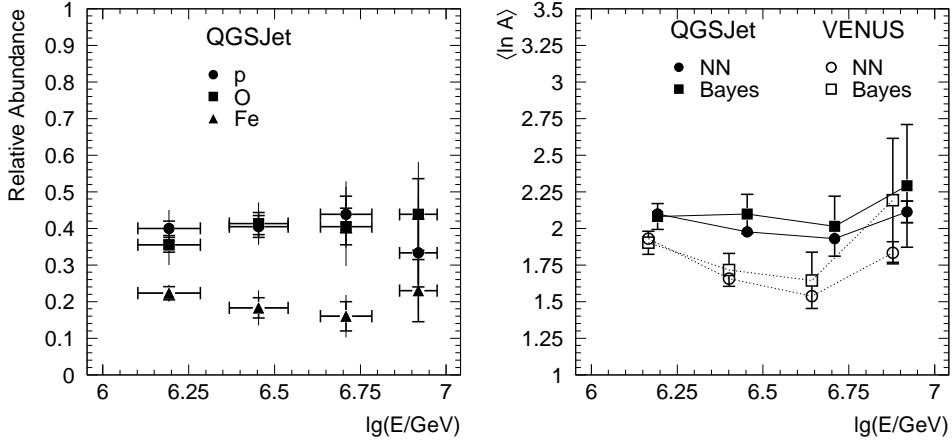,width=\textwidth}
  \caption{Relative abundances reconstructed by Bayes classification vs. the
    reconstructed  energy based on the QGSJet model and using $\Ne$ and
    $\Nmutr$. Additionally, the corresponding mean logarithmic mass $\langle\ln
    A\rangle$ (right; Bayes) and  the 
    corresponding variation resulting from the  neural network analysis (NN)
    are given. The error bars represent the  statistical (thick) and
    methodical (thin) uncertainties.} 
  \label{comp2meanenergy_paper:fig}
\end{figure}

As  an example,  Figure~\ref{comp2meanenergy_paper:fig} shows  the relative
abundances of a mass classification  into three categories, as well as the
corresponding mean logarithmic mass, resulting from the analysis of {\em
  central\/} showers (selection II) vs. the estimated energy. The observables
$\Ne$ and $\Nmutr$ are used as input parameters. 
Again, the VENUS model leads to a lighter mean logarithmic mass in the
considered energy range as compared to the QGSJet model.
Besides the Bayesian method a neural network analysis was performed 
additionally. The network results are denoted by NN. 
Within the statistical errors the mass composition resulting from both pattern
recognition procedures agree. Our data reveal a mixed composition, becoming
lighter when approaching the knee and heavier above the knee. This feature
appears to be somehow mysterious, but the tendency  is supported by recent
results of the CASA-BLANCA~\cite{CASA-BLANCA} and HEGRA~\cite{HEGRA}
experiments. In fact, there are also astrophysical arguments for  a minimum of
the mean mass in the range of the knee~\cite{swordy}. 

In the present status of our analysis procedure it is hardly possible to
introduce more than three classes for the reconstruction of the mass
composition. If this were to be attempted additional observables had to be
included. A finer binning of the energy
scale (beyond the energy resolution $(\Delta E/E)_{\rm est}$) for the spectra
of single masses would require to deconvolute the resolution effects. In the
actual analysis this step has not been performed and only a few representative
values of the varying mass composition (and no detailed energy spectra of the
different mass classes) have been presented. To analyze the data beyond this
limit we need, in the simplest case, to construct from the misclassification
matrices a matrix $A_{\rm AA';EE'}$ deconvoluting  mass and energy
resolution effects. Stressing once more the curse of dimensionality (see
Appendix~\ref{sec:methods}), a very large
number of simulated events is required for the determination of such a
matrix (at least 150000 simulated events are needed). For the same 
reasons we are presently unable to infer any significant 
fine structure from the all-particle energy spectrum beyond the resolution
$(\Delta E/E)_{\rm est}$. 

Different sets of observables lead to different mass compositions. This
feature may indicate that not only the absolute values, but also the
correlations of the observables are not described satisfactorily by any of
the models. Nonetheless, a reduced model dependence can be observed when a well
marked relation is projected out from the correlations of the multivariate
distribution.  
\begin{figure}[bt]
  \centering
  \epsfig{file=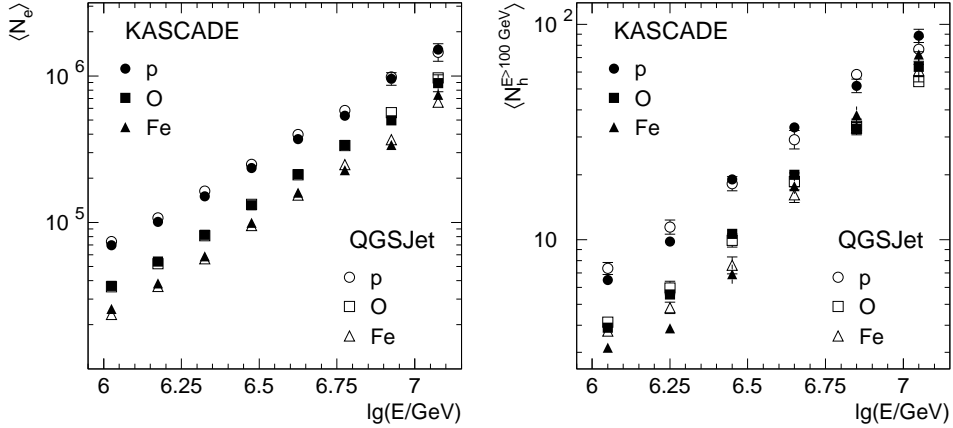,width=\textwidth}
  \caption{The projected relations $\lg \Ne$ and $\lg N_{\rm h}^{
      E>100\unit{GeV}}$, respectively, vs. $\lg(E/\unit{GeV})$
    from two neural networks,
    trained to estimate the energy and mass of the measured events using $\Ne$
    and $\Nmutr$ as EAS observables.} 
  \label{corr_paper2:fig}
\end{figure}
Figure~\ref{corr_paper2:fig} displays on the left-hand side the relation
between 
electron size $\Ne$ and energy $E$, which shows no difference between
simulated and measured samples. Of course, this is not surprising, since the
pattern recognition tool is just trained in that way, such that deviations,
incompatible with the statistical accuracy, would cast some methodical doubts
on the used algorithms. More remarkable is the agreement of
$N_{\rm h}^{E>\unit[100]{GeV}}$ vs. primary energy, found by the same
network though with larger fluctuations of the mean values. That may be
explained by the reduced mass sensitivity  of $N_{\rm h}^{
E>\unit[100]{GeV}}$ and the dominance of the $\Ne$-$\Nmutr$ correlation
(compare the observable sets in Figure~\ref{comptest:fig} left). 
Nevertheless, within the statistical significance
level (in terms of hypotheses tests like student $t$-test)  no difference
between data and model predictions can be stated.


\section{Discussion and conclusion}
The present paper aims at presenting methods of a determination of primary 
energy spectrum and mass composition of cosmic rays in the energy range
\mbox{$\unit[10^{15}-5\cdot10^{16}]{eV}$} by an event-by-event analysis
of EAS data. The specific methodical feature is the
use of a non-parametric approach, studying multivariate distributions of a
number of EAS observables~\cite{ashot90,ashot97}. 

The present approach to obtain information
about the EAS primaries has following merits: 
\begin{itemize}
\item It specifies the inevitable model dependence of any statement about
  spectrum and mass composition, introduced through the patterns provided by
  the Monte Carlo simulations on basis of a particular hadronic interaction
  model. 
\item The model dependence is not only revealed by the results from the 
  analysis  
  of single EAS observables when comparing different hadronic interaction
  models, but the approach specifies also the degree of correlations between
  different observables used for the multivariate analysis.  
\item This feature provides the possibility to test a specific hadronic
  interaction model by exploring the internal consistency of the results, when
  the outcome of different sets of observables are considered. This aspect is
  of greatest importance for approaching the best model
  reproducing the observations in the most consistent way.
\item Comparing the KASCADE findings with other experiments shows that 
the discrepancies between results can well be attributed to the different interaction models employed.
\end{itemize}

\ack
We acknowledge various clarifying discussions with Ralph Engel, Sergej
Ostap\-chenko and Klaus Werner about the use and embedding of the hadronic
interaction models Sibyll, QGSJet and VENUS in the Monte Carlo EAS simulation
code CORSIKA.  The authors would like to thank the members of the engineering 
and technical staff of the KASCADE collaboration who contributed with 
enthusiasm and engagement to the success of the experiment. 
The work has been supported by the Ministry for Research of the
Federal Government of Germany, by a grant of the Romanian
National  Agency for Science, Research and Technology, by a research
grant (No.~94964) of the Armenian Government,
 and by the ISTC project
A116. The collaborating group of the Cosmic Ray Division of the Soltan
Institute of Nuclear Studies in Lodz and of the University of Lodz is
supported by the Polish State Committee for Scientific Research. The KASCADE
collaboration work is embedded in the frame of scientific-technical
cooperation (WTZ) projects between  Germany and Romania (No.~RUM-014-97),
Poland (No.~POL~99/005) and Armenia  (No.~002-98).
%


\newpage
\begin{appendix}
\section{Non-parametric statistical inference}
\label{sec:methods}

Pattern recognition techniques are efficient tools to determine the correct
association of a given sample to a certain category or class.
From the measurements or simulations of a physical phenomenon, a set of
quantities (observables) is obtained, like $\Nmutr$ or $\Ne$, which defines an
observation vector $x$. This observation vector serves as the input to a
procedure based on decision rules, by which a sample  is assigned to one of
the given classes. Thus it is assumed that an observation vector is a random
vector $x$ whose conditional density function $p(x|\omega_i
)$ depends on its  class $\omega_i$
(e.g. p, O and Fe classes).  

In the following we consider so called {\em non-parametric\/} techniques like
Bayes classifiers and artificial neural networks~\cite{ashot89}. The term
non-parametric 
indicates that the representations of the distributions (like probability
density functions of Bayes classifiers or weights of neural networks) are no
more  specified by a-priori chosen functional forms. They are constructed
through the analysis process by the given data distributions themselves. 

It should be immediately emphasized that there are  some  important
limitations. In case of a finite set of 
random samples, the dimension $n$ of the random vector $x$  is limited by the 
following condition:
When considering each component of an $n$-dimensional  observation vector by
$M$ divisions, 
the total number of cells is $M^n$ and is increasing exponentially with the
dimensionality of the input space. Since each cell should contain at least one
data point this requirement implies that the size of training samples (or
reference pattern samples) needed to specify the non-parametric mapping, is
increasing correspondingly. This condition is called the {\em curse of
 dimensionality}~\cite{bishop95} and prohibits the simultaneous
(multivariate) analysis of a larger number of EAS observables, when the
size of training samples is too small. 

\subsection{Bayesian decision rule}
\label{sec:bayes}
The Bayes classifier is a powerful algorithm but time
consuming with large memory requirements. However, its performance is
generally excellent and asymptotically 
{\em Bayes optimal}, so that the expected {\em Bayes error} (see below) is
less than or equal to  that of any other technique~\cite{fukunaka72}. The
estimated probability densities converge asymptotically to the true density
with increasing  sample size~\cite{parzen62,cacoullos66}. 

\begin{figure}[t]
\epsfig{file=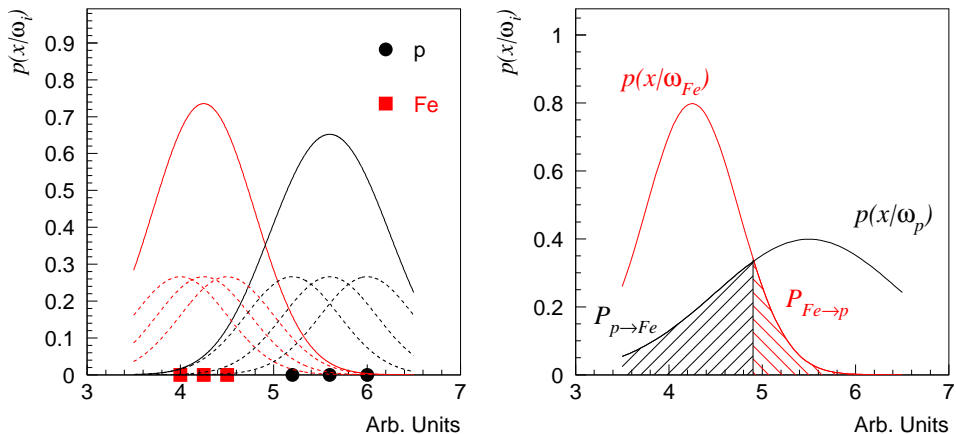,width=\textwidth}
  \caption{Schematic illustration of the construction  of two one-dimensional
    (overlapping)  
    likelihood functions $p(x|\omega_{\rm p,Fe})$, approximated by Gaussian
    distributions ({\em sphere-of-influence}) for each event, indicated on the
    abscissa (left). Classification using the Bayes decision showing the  
    proportion of mis-classified events by the hashed areas (right)
    ($P(\omega_{Fe})=P(\omega_p))$ .}
           \label{callgaussbayes_paper:fig}
\end{figure}

The method is based on the {\em Bayes Theorem}~\cite{bayes1763}
\begin{equation}
p(\omega_i|x)=\frac{p(x|\omega_i)\times P(\omega_i)}{p(x)} \quad 
\Leftrightarrow \quad \mbox{posterior} = 
\frac{\mbox{likelihood}\times \mbox{prior}}{\mbox{normalization factor}}
\label{eqn:bayestheorem}
\end{equation}
with $p(x)=\sum_{j=1}^N p(x|\omega_j)P(\omega_j)$, which holds if the 
different $N$ hypotheses $\omega_i$ (i.e. classes) are mutually exclusive and 
exhaustive. By a  prior and a normalization factor the theorem connects the 
likelihood for an event $x$ of a given class $\omega_i$ with the 
probability of a class $\omega_i$, being associated to a given event $x$.
The prior gives the {\em a priori} knowledge of the relative abundance of each
class and is major basis of debates on Bayesian inference procedures. It is
nearly always the best to follow the advice given by Bayes
himself~\cite{bayes1763}, generally known as  
{\em Bayes' Postulate} (occasionally also referred to as 
{\em Principle of Equidistribution of Ignorance}): 
So far there exists no further knowledge, the prior probabilities 
should be assumed to be equal
\begin{equation}
P(\omega_i)=\frac{1}{N}  \quad\mbox{ with }\quad  \sum_{i=1}^N P(\omega_i)=1
\label{eqn:prior}
\end{equation}
In the fortuitous case that the likelihood functions $p(x|\omega_i)$ 
are known for all populations, the Bayes optimal decision rule is to classify 
$x$ into class $\omega_i$, if 
\begin{equation}
p(\omega_i|x)>p(\omega_j|x) 
\label{eqn:bayesrule}
\end{equation}
for all classes $\omega_j\neq \omega_i$, as illustrated (with the
mis-classification probabilities) in
Figure~\ref{callgaussbayes_paper:fig}. 

To construct an estimate $\hat{p}(x|\omega_i)$ of the likelihood
$p(x|\omega_i)$ of class $\omega_i$, the $k$-th simulated event $x_{ki}$  is
assumed to have a {\em sphere-of-influence} where it contributes to the
probabilities (see Figure~\ref{callgaussbayes_paper:fig}).
There are various procedures to specify these contributions whose
superpositions lead to continuous likelihood functions, replacing the
frequency distributions of discrete simulated events $N_i$ of each class
$\omega_i$ in the $n$-dimensional {\em  observation} space.
A  standard choice of such {\em spheres} are multivariate normal
distributions, because they are simple, well behaved, easily computed and have
been shown in practice to perform well: 
\begin{equation}
\hat{p}(x|\omega_i)=\frac{1}{N_i}\sum_{k=1}^{N_i}\frac{1}{(\sqrt{2\pi}\sigma)
^{n+1} \sqrt{|C_i|}}
\exp\left(-\frac{1}{2 \sigma^2}||x-x_{ki}||\right)
\label{eqn:gauss}
\end{equation}
The {\em Mahalanobis metric}  $||x-x_{ki}||=(x-x_{ki})^TC_i^{-1}(x-x_{ki})$ is 
used, because the observables 
are transformed 
to unit variances by the sampling covariance matrix $C_i$ for each class 
$\omega_i$, resulting in equal importance of all components. 

The scaling parameter $\sigma$ controls the width of the sphere-of-influence
and is obtained by the median of an ordered statistics of estimated {\em Bayes
  errors} for different initial values of $\sigma$~\cite{chiliguide}.  
The {\em Bayes error} $\epsilon$ represents the total sum (integral) of 
mis-classified events and is given in case of two classes by the simple
relation 
$
\epsilon=\int \min\{p(\omega_1|x),p(\omega_2|x)\}\cdot p(x) \d x
$ (hatched areas in Figure~\ref{callgaussbayes_paper:fig} right).
To account for the mis-classification, the rates 
$P_{ij}=P_{\omega_i \rightarrow \omega_j}$,
i.e. the probability of an event $x\in \omega_i$ being classified in the class
$\omega_j$, are estimated by the {\em leave-one-out} method (also called {\em
  jack-knifing}). Each simulated event is held back once while the others are
used to estimate the association of this particular event.
By a, so called, {\em bootstrap} method
different  subsets of each simulated class are used to perform the {\em
  leave-one-out} method to give an asymptotically unbiased estimate of the
variance of the $\hat{P}_{ij}$~\cite{fukunaka72}. 
Thus the true number of events $n_i^\star$ can be deduced from the classified
events $n_j$ by a matrix inversion: 
\begin{equation}
\label{eqn:reconstr}
\sum_j \hat{P}_{ij}^{-1}n_j=n_i^\star \qquad\mbox{ with } \quad
  \hat{P}_{ij}=\hat{P}_{\omega_i\rightarrow 
  \omega_j}
\end{equation}
\subsection{Neural networks}
\label{sec:neural}
An artificial neural network can be considered as a nonlinear transfer
function  
\begin{equation}
f:\Rset^p\longrightarrow \Rset^q
\end{equation}
mapping a bounded euclidian space of dimension $p$ to another space of
dimension $q$. 
The, so called, {\em multilayer feed-forward} neural network is organized in
$L$ different layers: an input layer, 
$L-2$ hidden layers, and an output layer. 
Each layer $l$ consists of a certain 
number $n_l$ of units (neurons), which carry on the {\em signals} to the next
layer. The, so called, {\em network topology} specifies the number of units in
each layer:  
$n_1 \times n_2 \times \ldots \times n_{L-1} \times n_L$.
An {\em output unit} $y_{mL}$ of the output layer $L$ is determined for each 
observation vector ({\em input units}) $x_{ki}$ and 
class $\omega_i$ entering the input layer and should be close to the true 
value $t_{ki}$, given by the labeled simulation events in terms of a well 
defined measure.
Thus, the error function $E({\bf w})$
\begin{equation}
\label{eqn:nngoal}
E({\bf w})=\frac{1}{2} \sum_{i=1}^N \frac{1}{N_i} \sum_{k=1}^{N_i} 
\left( y_{mL}(x_{ki},{\bf w})-t_{ki}\right)^2
\end{equation}
has to be minimized.
For each layer $l$, except of the input, the outcome of each neuron $m$ is
calculated by a weighted sum of the output of neurons of  
the last preceding layer. Additionally an {\em activation function} $f(z)$ is
applied to the sum
\begin{equation}
y_{ml}=f(z)=f\left(\sum_{i=1}^{n_{l-1}} 
                        w_{i,l-1}^{m}\cdot y_{i,l-1} +w^{m}_{l}\right).
\end{equation}
A convenient practice is to use the Fermi function $f(z)=1/(1+\exp(-z))$.
The most common algorithm for the network training, i.e. minimizing
$E({\bf w})$,   
is an adjustment of the weights $w_{i,l-1}^{m}$ and $w^{m}_{l}$
by a stochastic minimization procedure~\cite{chiliguide}  or alternatively
by the, so called, {\em back-propagation} algorithm~\cite{rummelhart86}.
There exist different other algorithms or extended versions of this basic
back-propagation, which  try to circumvent problems in finding the global
minimum or sticking in a  
local minimum. Additional problems arise, if the training process leads to an
overtraining of the network by adopting the properties of the training
samples, but cannot give satisfactory results, when it is applied to
another validation 
set. Thus, in a generalization phase one has to control the quality of the
network with an independent labeled set of samples.
 
In general, the output $y_{mL}$ is a continuous function. Hence  not only the
classification can be done applying neural networks, but also parameter 
estimation (regression) is possible, e.g. the estimation of the primary energy 
of EAS events.
In case a classification into $N$ classes is required, the output
$y_{mL}$ of the network is divided into $N$ regions, each representing a 
single class~\cite{chiliguide}.

In previous publications the consistency and equivalence
of neural network and Bayes classifier results in EAS analysis 
have been demonstrated~\cite{roth99,ashot97}. 
The classification rates $P_{ij}$  inferred from 
both procedures do not differ significantly. Thus, 
an adequate choice of the particular decision rule and of the appropriate
algorithm is just a matter of the actual conditions like computing time  
and memory workload.


\end{appendix}

\end{document}